\documentclass[letterpaper]{article}

\usepackage[T1]{fontenc}

\usepackage{geometry}
\usepackage{setspace}
\usepackage{wrapfig}
\usepackage[style = chem-acs]{biblatex}
\usepackage{booktabs}
\usepackage{graphicx}
\usepackage{float}
\newfloat{scheme}{htbp}{los}
\floatname{scheme}{Scheme}
\floatname{chart}{Chart}
\newfloat{graph}{htbp}{loh}

\usepackage{chemformula} 
\usepackage[version = 4]{mhchem} 

\usepackage{authblk}
\author[1,2]{Max C. Gallant}
\author[1]{David Mrdjenovich}
\author[1,2]{Kristin A. Persson*}
\affil[1]{Department of Materials Science and Engineering, University of California, Berkeley, CA}
\affil[2]{Materials Sciences Division, Lawrence Berkeley National Laboratory, Berkeley, CA}

\title{An Investigation in the Kinetic Persistence of TiO$_2$ Polymorphs using Machine Learning Driven Pathfinding in Crystal Configuration Space}
\date{*Email: kristinpersson@berkeley.edu}

\begin{document}

\maketitle

\begin{abstract}
As the number of theoretically predicted materials continues to grow, it becomes increasingly important to assess not only their thermodynamic stability but also their kinetic viability under realistic synthesis conditions. In this study, we investigate the hypothesis that the kinetic persistence of a metastable polymorph is related to the topography of the potential energy landscape separating it from lower energy phases. To accomplish this, we develop a new method for identifying diffusionless transformation pathways between metastable polymorphs and their ground-state counterparts and discuss the energetics of those pathways with respect to the experimental observation of each phase. This algorithm is underpinned by the recently developed Crystal Normal Form, which provides a graph representation of crystal configuration space and supplies the substrate for our pathfinding algorithm. We apply this method to the \ce{TiO2} system which contains the well-known anatase, rutile, and brookite phases in addition to a number of hypothetical metastable polymorphs.
\end{abstract}

\section*{Keywords}

synthesizability, metastable polymorphs, diffusionless transformation, machine learning potential, materials science, crystal structure, transformation, pathfinding

\section{Introduction}

\label{sec:introduction}

Over the past decade, the field of solid-state synthesis science, and in particular its theory and modeling aspects, has received increased attention due to its central role in developing novel inorganic functional materials. The relative recency of this focus is related to the dual rise of high-throughput simulation methods for determining material properties and the adoption of machine learning methods for computational materials science. Specifically, the data generated by high-throughput simulation have provided the fuel for the application of advanced generative deep learning methods to propose hypothetical new crystal structures \cite{xie_crystal_2021, zeni_generative_2025, luo_crystalflow_2025}. These methods have made it relatively straightforward to generate proposals for crystal structures that satisfy a set of pre-defined performance targets. The resulting accumulation of unverified candidate structures has given rise to the phrase "synthesis bottleneck" which refers to fact that successful synthesis of such hypothetical crystal structures lags behind. 

In response to this challenge, significant effort has been made to develop methods to relieve the bottleneck by aiding in solid-state synthesis route design. Among these methods are extensive natural language processing campaigns on published synthesis literature \cite{kononova_text-mined_2019, cruse_text_2024, he_similarity_2020, zhang_named_2024}, metrics of synthesizability based on thermodynamics \cite{mcdermott_assessing_2023, szymanski_quantifying_2024, sun_thermodynamic_2016, aykol_thermodynamic_2018, tolborg_free_2022}, algorithms for predicting synthesis outcomes \cite{mcdermott_graph-based_2021, gallant_cellular_2025, karan_ion_2025}, and automated methods for designing synthesis routes \cite{he_precursor_2023, szymanski_autonomous_2023, chen_navigating_2024}. Many of these approaches rely on thermodynamic features such as formation energies or phase diagram and convex hull geometry. While these features contain information about synthesizability \cite{aykol_thermodynamic_2018}, thermodynamics alone do not provide sufficient predictive power to determine whether or not a metastable polymorph is kinetically stabilized as compared to other, more stable, polymorphs.\cite{sun_thermodynamic_2016}  Furthermore, while dynamically unstable polymorphs can be excluded through full phonon calculations, this criterion alone does not guarantee the elimination of all kinetically non-persistent structures. Even in well-studied binary systems such as \ce{Fe2O3} and \ce{TiO2}, a substantial fraction of unobserved, low-energy hypothetical polymorphs exhibit no imaginary phonon modes, indicating true metastability.\cite{sun_thermodynamic_2016} This raises a fundamental question: why do many polymorphs that are both thermodynamically viable and dynamically stable remain experimentally unrealized?

In previous investigations of the impact of kinetics on synthesis outcomes, Karan et al. identified diffusion in amorphous intermediates as a descriptor of product selectivity \cite{karan_ion_2025} and Murat et al. developed a metric based on structural similarity to predict the ease of nucleation of a product phase \cite{aykol_rational_2021}. Additionally, several methods use the topography of the potential energy landscape (PES) to determine whether or not a given polymorph is synthesizable. One example of this approach connects the "basin hypervolume", or the extent of the region of configuration space for which a given polymorph is an attractor, to the synthesizability of that polymorph \cite{novick_basin-size_2025, stevanovic_sampling_2016}. Another approach correlates the synthesizability of a polymorph to the absence of a facile transformation pathway from that phase to a lower energy configuration \cite{stevanovic_predicting_2018}. It is this mode of kinetic persistence that we investigate in this study.

Although phase transitions in periodic solids (e.g. martensitic transformations) are more often mediated by nucleation and growth than by diffusionless transformations, previous work has illustrated that the energy barriers associated with such transformations can yield insights regarding the kinetic stabilization of polymorphs \cite{stevanovic_predicting_2018, schon_studying_1996, therrien_metastable_2021}. The challenge associated with identifying energy barriers for such transformations is in the problem of exploring the PES to identify the transformation pathway. Methods for addressing this problem range from sampling methods such as transition path sampling \cite{zhu_phase_2019} and stochastic surface walking (SSW) \cite{shang_stochastic_2013} to methods based on structure matching and symmetry \cite{wang_crystal-structure_2024, capillas_maximal_2007}.

Among these approaches, the nudged elastic band (NEB) method \cite{jnsson_nudged_1998} and its extensions, climbing-image NEB (CI-NEB) \cite{henkelman_climbing_2000}, solid-state NEB (ssNEB) \cite{sheppard_generalized_2012}, and variable-cell NEB (VC-NEB) \cite{qian_variable_2013}, stand out as the most widely used for identifying solid–solid transition pathways \cite{chakraborty_martensitic_2015, zarkevich_ferh_2018, eymeoud_impact_2021, trinkle_new_2003}. In this method, a number of intermediate configurations (images) are jointly relaxed in the presence of an added artificial spring force which keeps them roughly evenly spread out in configuration space along the reaction coordinate. NEB and its variants are  effective but rely on the provisioning of an initial pathway guess. Stevanovi\'c et al. address this issue with an algorithm for identifying the unit cell correspondence between endpoint structures that minimizes the changes in coordination number (\# of broken bonds) along the pathway. They show that this method yields high quality, low energy pathways between a number of polymorph pairs and provides evidence for the hypothesis that metastable polymorphs which are connected to a lower energy structure via a facile diffusionless transition pathway are less frequently observed in experiment \cite{stevanovic_predicting_2018}.

Here, we continue investigating this hypothesis and introduce an algorithm that identifies low energy pathways between polymorphs without relying on any prior information regarding the nature of the transition. This method is based on the Crystal Normal Form, which provides a graph representation of crystal configuration space, resolves the ambiguity of unit cell choice \cite{mrdjenovich_crystallographic_2024}, and enables the traversal of crystallographic phase space without the risk of revisiting the same structure. In their original work, Mrdjenovich and Persson utilize a water-filling algorithm which exhaustively searches the phase space graph to identify the lowest energy transition point separating two polymorph endpoints. However, this exhaustive approach is only applicable to simple structures and suffers from combinatorially growing computational requirements as the number of atoms in the unit cell increases.  Here, we combine the CNF graph with an algorithm for identifying low-energy transition pathways and apply the combined framework to the \ce{TiO2} system. As previously noted, the \ce{TiO2} system presents an excellent test case as it contains several technologically important polymorphs (rutile, anatase, and brookite), and a number of experimentally unobserved (``theoretical''), dynamically stable, polymorphs within the energetic spectrum of the observed phases.\cite{sun_thermodynamic_2016} We identify diffusionless transformation pathways and energy barriers separating the theoretical polymorphs and the known polymorphs and correlate the height of these energy barriers to experimental observations.

\section{Description of the pathfinding algorithm}

\begin{figure}
    \centering
    \includegraphics[width=\linewidth]{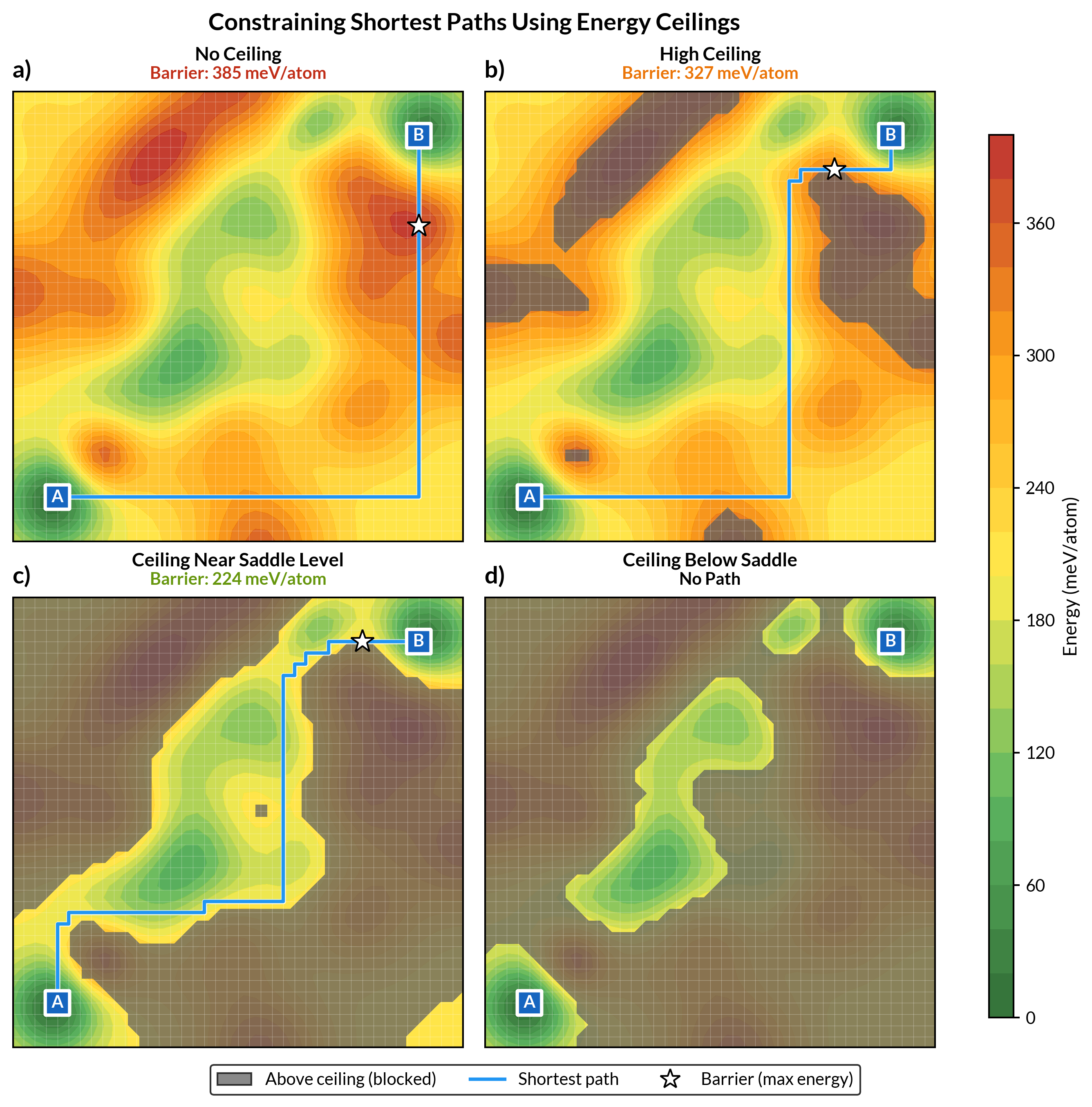}
    \caption{\textbf{Ceiling lowering algorithm illustrated on a schematic energy landscape.} The topography of the landscape in this figure was procedurally generated for illustration purposes but does not correspond to any real potential energy surface. The two square endpoints at local minima (A, B) illustrate how metastable polymorphs sit at local minima on the surface. In panels a)-d), grey areas correspond to configurations with energies above the energy ceiling for that panel. The blue paths are produced using the A* shortest-path finding algorithm on the accessible region of each space while only allowing steps to immediate neighboring squares. a) Without any energy ceiling limit imposed, the shortest path traverses an energy maximum, leading to a high ``barrier''.  b), c) As the ceiling approaches the highest saddle separating the two endpoints, the shortest path gets closer to traversing the true saddle point. d) If the ceiling is lower than the true saddle point, no percolating pathway exists, and not transformation pathway is found. The stars mark the highest energy points encountered on each pathway.}
    \label{fig:ceiling_algo}
\end{figure}

The Crystal Normal Form (CNF) presents an ordered list of integers which uniquely describes a crystal structure and resolves the ambiguities associated with the conventional unit cell representation. In the CNF, the lattice is represented by the lengths of seven ``Voronoi vectors'' (lattice generating vectors whose bisecting planes intersect the Wigner-Seitz cell) and the motif is represented by a list of relative atomic coordinates. By construction, if two unit cells (lattice/motif pairs) generate the same crystal structure, their CNF representations will be identical. In this sense, the CNF acts as a fingerprint for a crystal structure and as a method of identifying distinct structures.

In addition to this capacity for disambiguation, Mrdjenovich et al. developed a method for enumerating a set of neighbors for a given CNF point. These neighboring CNF points correspond to structures that are related to the original structure by either a small strain or phonon amplitude. By repeatedly identifying neighboring structures (and neighbors of neighbors), this method enables an exhaustive enumeration of all crystal configurations and produces a graph (``the CNF graph'') within which similar configurations are connected. Paths between points in this graph correspond to transition pathways composed of small strains or phonon amplitudes connecting polymorphs. These paths hold tremendous expressive power because these lattice and motif steps can be interleaved in whatever way is necessary to connect the endpoints. The magnitude of the structural differences between neighboring points in the graph are determined by two discretization parameters, $\xi$ and $\delta$ which determine the sizes of the lattice strain steps and atomic displacement steps respectively.

As previously mentioned, for the simplest unit cells ($< 2$ atoms per cell), it is possible to exhaustively search this CNF graph using a water-filling algorithm in conjunction with density functional theory (DFT) calculations for assessing the energy of each point to first-principles accuracy. In this water-filling algorithm, a starting configuration is identified and then neighbors are explored in order of ascending energy such that no configuration is visited until all lower-energy configurations have been explored. When this algorithm reaches a saddle point, it will repeatedly explore the lowest energy point on the opposite side of the saddle until the local minimum for that basin is reached. If that basin is the endpoint, the algorithm will terminate. If that basin is not the endpoint, the algorithm will proceed as before: by exhaustively exploring points in order of ascending energy until yet another saddle point is reached and the algorithm search frontier (the ``water''), expands (``flows'') down into the next basin.

Such an approach is attractive because it is guaranteed to find the lowest energy barrier separating two polymorphs (up to the discretization parameters), but it quickly becomes infeasible due to the high compute cost of DFT calculations. This limitation can be partially mitigated by replacing DFT with an MLIP surrogate; however, the search space still grows combinatorially with system size, restricting practical applications to only modestly larger unit cells (typically $<4–6$ atoms, depending on the complexity of the problem). An alternative method uses the A* shortest-pathfinding algorithm \cite{hart_formal_1968} in conjunction with a Manhattan distance heuristic on the CNF coordinate list to quickly (CPU seconds/minutes) find \textit{short} paths between polymorphs. While this approach expands the range of structures for which transition pathways can be identified, it does not account for energetics and therefore does not guarantee low-energy pathways.

\begin{figure}[ht]
    \centering
    \includegraphics[width=\linewidth]{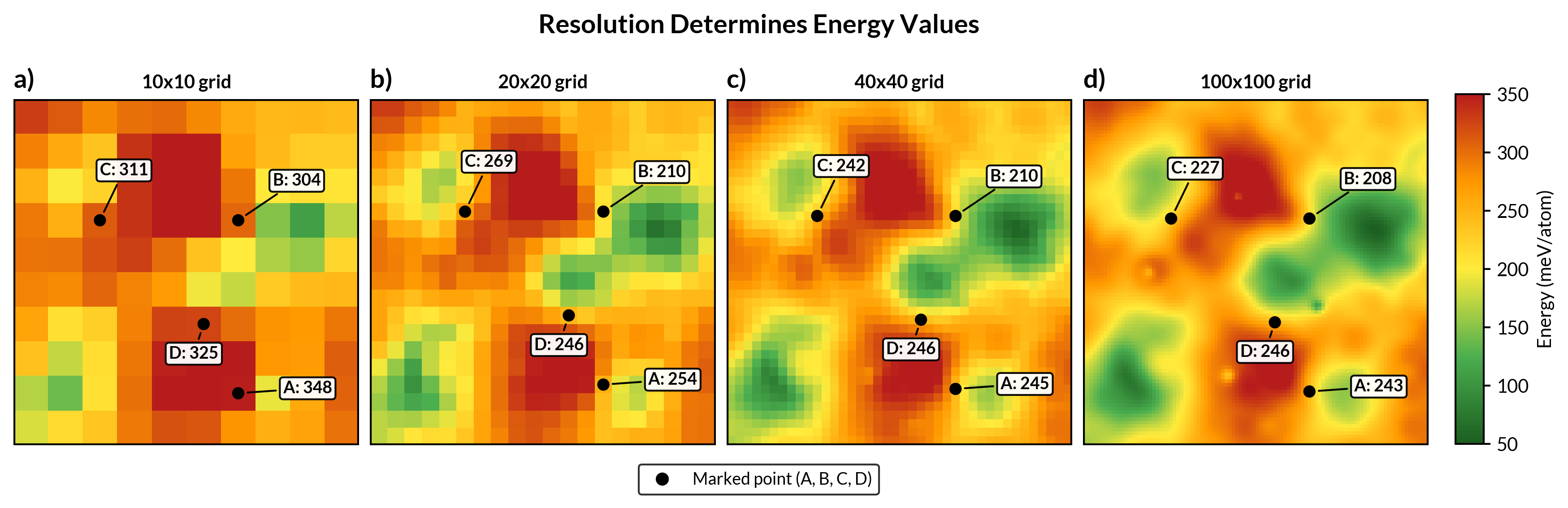}
    \caption{\textbf{Resolution affects landscape features.} Panels a) through d) show increasingly high resolution views of the same schematic potential energy landscape. At lower resolutions, fine details are not visible. Because configurations snap to the point nearest them under the CNF discretization, the energy of a point depends on the discretization parameters chosen. This is illustrated here by the changing energy values associated with points labeled A, B, C, and D under different discretization levels.}
    \label{fig:discretization}
\end{figure}

To combine the benefits of both approaches (that is, to find low-energy paths between larger unit cells), we introduce an iterative refinement algorithm that applies A* pathfinding to progressively smaller subgraphs of the complete CNF graph by repeated lowering of an energy ``ceiling''. The energy ceiling is a filter applied to the CNF graph which excludes nodes that have energies higher than the ceiling from exploration. Different energy ceilings leave different portions of the CNF graph accessible to the pathfinding algorithm. Specifically, for a given ceiling energy, $C$, any path found must only visit nodes with a maximum energy value $C$. Once a path has been found under $C$ achieving a maximum energy of $E_{max} < C$, the ceiling can be lowered to $E_{max} - \delta C$ (with $\delta C = 2 \frac{meV}{atom}$, for example - though this is a tunable parameter), and the pathfinding process can be repeated. Thus, with each subsequent lowering of the ceiling and successful pathfinding attempt a lower energy transition pathway is identified. A schematic showing how the shortest path varies with energy ceiling level on an artificial 2D potential energy landscape (generated procedurally for illustration purposes only) is shown in Figure \ref{fig:ceiling_algo}.

It is important to remember that the CNF energy landscape is a discretized space. As a result, features of the landscape may appear or disappear as the coarseness of the discretization is adjusted. Typically, finer low energy features (for example, narrow saddle points or low energy valleys) become more accessible at higher resolutions, as shown schematically in Figure \ref{fig:discretization}.If, after several failed pathfinding attempts at a given energy ceiling, no path is found, the resolution of the search space can be increased to reveal finer details of the energy surface and a new round of attempts can commence. Ultimately, after a prescribed number of iterations, the algorithm terminates, and the lowest ceiling under which a pathway was found is treated as an upper bound on the true pathway barrier.

\section{Results and Discussion}

At present, the Materials Project reports 46 polymorphs of \ce{TiO2} \cite{horton_accelerated_2025}, however, recognizing thermodynamic constraints, we restrict this study to polymorphs within a defined energy-above-hull window relative to the three most common naturally occurring phases: anatase, brookite, and rutile. This filtering procedure yields 10 polymorphs including the three most well-known experimental phases and seven other polymorphs (including columbite, baddeleyite and $\beta$-\ce{TiO2}, which have been reported in high-pressure experiments as shown in Table \ref{tab:polymorphs}). We filtered this set to polymorphs with 24 or fewer atoms (shown in Table \ref{tab:polymorphs}), which led to the exclusion of the orthorhombic mp-775938 phase (36 atoms/primitive cell). We also ultimately excluded mp-34688 from the polymorph list because the CNF revealed that this structure constitutes a minor distortion of anatase. Specifically, the transformation paths connecting this phase to anatase were < 10 steps long even at a fine discretization. This similarity was confirmed by the structure matching algorithm in pymatgen \cite{ong_python_2013} which reported a match and the Earth Mover's Distance on the Pointwise Distance Distribution \cite{widdowson_pointwise_2025} representation (0.13 \AA).

\begin{table}[htbp]
  \centering
  \begin{tabular}{llccccc}
  \toprule
  MP ID & Phase Name & Other Names & Space Group & Atoms & $E_\mathrm{hull}$ (meV/atom) & Status \\
  \midrule
  \multicolumn{7}{l}{\textit{Observed (ambient pressure)}} \\
  mp-390 & Anatase & --- & $I4_1/amd$ & 6 & 0 & Observed \\
  mp-1840 & Brookite & --- & $Pbca$ & 24 & 3 & Observed \\
  mp-2657 & Rutile & --- & $P4_2/mnm$ & 12 & 44 & Observed \\
  \midrule
  \multicolumn{7}{l}{\textit{Observed (high pressure)}} \\
  mp-1439 & Columbite & $\alpha$-\ce{TiO2}, \ce{TiO2}-II & $Pbcn$ & 12 & 5 & Observed \\
  mp-554278 & $\beta$-\ce{TiO2} & \ce{TiO2}-B & $C2/m$ & 12 & 6 & Observed \\
  mp-430 & Baddeleyite & --- & $P2_1/c$ & 12 & 39 & Observed \\
  \midrule
  \multicolumn{7}{l}{\textit{Theoretical}} \\
  mp-34688 & --- & --- & $C2/c$ & 6 & 15 & Theoretical \\
  mp-2420244 & --- & --- & $Pnma\text{-}I$ & 12 & 16 & Theoretical \\
  mp-754769 & --- & --- & $Pnma\text{-}II$ & 24 & 36 & Theoretical \\
  \bottomrule
  \end{tabular}
  \caption{\ce{TiO2} polymorphs selected for this study. Experimental polymorphs are the three well-known ambient-pressure phases. ``Observed'' indicates structures with entries in the Inorganic Crystal Structure Database (ICSD), typically synthesized under high-pressure conditions. Theoretical structures have no experimental reports in the ICSD.}
  \label{tab:polymorphs}
\end{table}

Any discussion of the energetics of \ce{TiO2} polymorphs should address the well-known variations in the energetic ordering of these polymorphs computed using various DFT methods \cite{luo_phase_2016, muscat_first-principles_2002, zhu_stability_2014}. Specifically, under the generalized gradient approximation\cite{perdew_generalized_1996} (GGA), anatase is identified as the ground state and brookite is lower in energy than rutile. Interestingly, more advanced methods, including diffusion Monte Carlo and finite-temperature approaches, predict anatase as the ground state at 0~K and rutile as the stable phase at elevated temperatures, while brookite remains higher in energy than both. Notably, all three polymorphs are separated by only small energy differences, spanning approximately 14meV/atom \cite{luo_phase_2016}. This close competition between rutile, brookite, and anatase is reflected in reports of successful synthesis and persistence of all three phases under ambient conditions \cite{castro_synthesis_2008, chen_new_2014, lee_synthesis_2006}.

Because we utilize MLIPs trained and finetuned on GGA calculations for this work, the barrier heights we compute are subject to a similar degree of error reflected in the incorrect ordering of these phases. We take this uncertainty to be roughly 50 meV/atom because this is the energy range spanned entirely by the anatase, brookite and rutile phases (wherein the ordering is incorrect) under GGA. As a result, and particularly given the additional uncertainty associated with using a finetuned MLIP to compute these pathways, we interpret our results qualitatively rather than as precise quantitative estimates of barrier heights.

\begin{figure}
    \centering
    \includegraphics[width=0.8\linewidth]{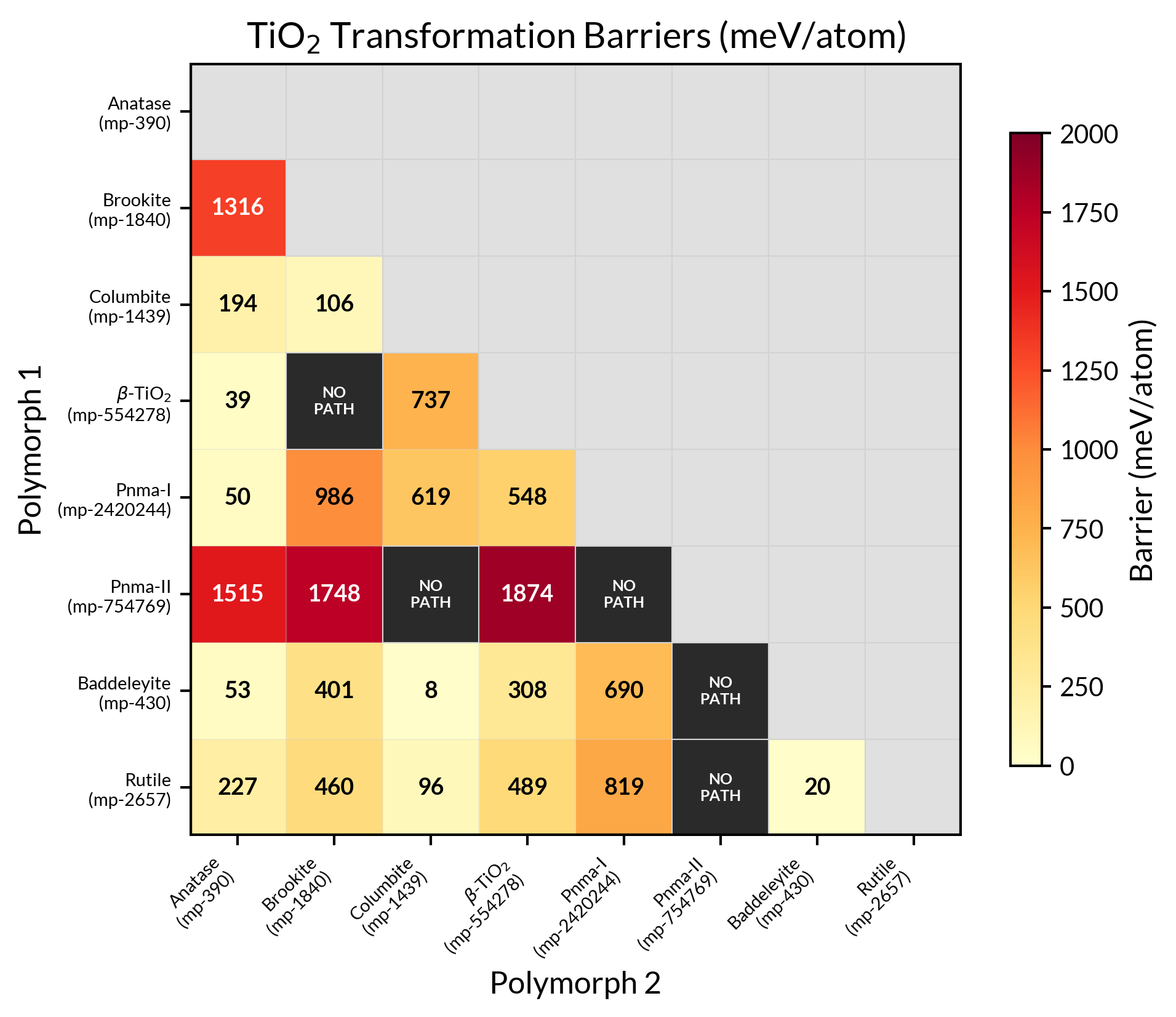}
    \caption{\textbf{Energy barriers between all pairs of polymorphs considered in meV/atom.} All values are in units of meV/atom.}
    \label{fig:barrier_matrix}
\end{figure}

We report the energy barriers identified by our algorithm in Figure \ref{fig:barrier_matrix}. All values are reported in meV/atom, but we note that this choice of unit suggests an intensive nature to the value. In fact, the energy required to activate a diffusionless transformation is an extensive property and these values should be regarded simply as descriptors for the difficulty of transformation. Additionally, it is important to keep in mind that due to the inaccuracies of DFT in ranking these phases energetically, we report barrier heights but intentionally do not specify a direction for the transformation. 

Our algorithm finds low-barrier (39-53 meV/atom) paths connecting $\beta$-\ce{TiO2}(mp-554278, hp), Pnma-I (mp-2420244, theoretical), and baddeleyite (mp-430, hp) to the anatase (mp-390) phase, suggesting that there is no significant kinetic barrier protecting these phases from decomposing to the naturally occurring anatase configuration. This result is consistent with the absence of observation of these phases under ambient conditions: baddeleyite (mp-430, hp) and $\beta$-\ce{TiO2} (mp-554278, hp) are known to occur only at high pressure, and no experimental record of Pnma-I (mp-2420244, theoretical) exists. Among these, the low energy (39 meV/atom) $\beta$-\ce{TiO2}$\leftrightarrow$anatase pathway finding is consistent with the work of Lu et al. \cite{lu_phase_2025} who found a facile pathway connecting the paths with a barrier height of 93meV/atom. The 227 meV/atom barrier height we find separating rutile (mp-430) and anatase (mp-390) is comparable to previously reported barriers for that transformation (233-333 meV/atom \cite{song_phase_2020, lu_phase_2025} from Lu et al. and Song et al.) as well. 

In addition to these moderate/low energy pathways, we find an extremely facile pathway between columbite and baddeleyite (8 meV/atom) which is in agreement with the finding of Wang et al. \cite{wang_theoretical_2021} that there is no substantive energy barrier inhibiting this transition. We also find a facile pathway (barrier height 20meV/atom) connecting baddeleyite (mp-430, hp) to rutile (mp-2657). This result further confirms the findings of Wang et al. \cite{wang_theoretical_2021}, who identified paths with barrier heights between 16-80meV/atom for the baddeleyite$\leftrightarrow$rutile transformation. We do note though that pathways in their study were computed using compressed unit cells and therefore are not directly comparable.

A previous study by Chen et al. investigated the transformation pathway between brookite and rutile \cite{chen_phase-transition_2022} and found that the mechanism was composed of a transformation from brookite to columbite followed by a  transformation from columbite to rutile. Our results are consistent with this finding: we find low energy barriers separating brookite and columbite (106 meV/atom) and columbite and rutile (96 meV/atom) and a higher energy barrier preventing brookite from directly decomposing to rutile (460 meV/atom). The barrier height we find for the rutile$\leftrightarrow$columbite transformation (96meV/atom) also agrees well with the findings of a previous study on the same pathway which identified a barrier height of 77meV/atom \cite{wang_theoretical_2020}. Finally, an important finding of the Chen et al. study was that they identified anatase as a "byproduct" of the brookite$\rightarrow$columbite$\rightarrow$rutile transformation pathway. Our results are also consistent with this hypothesis because we identify a moderately low barrier separating columbite and anatase (194meV/atom) suggesting that some of the intermediate columbite may be able to transform to anatase (as a byproduct) instead of rutile during decomposition.

\begin{figure}
    \centering
    \includegraphics[width=\linewidth]{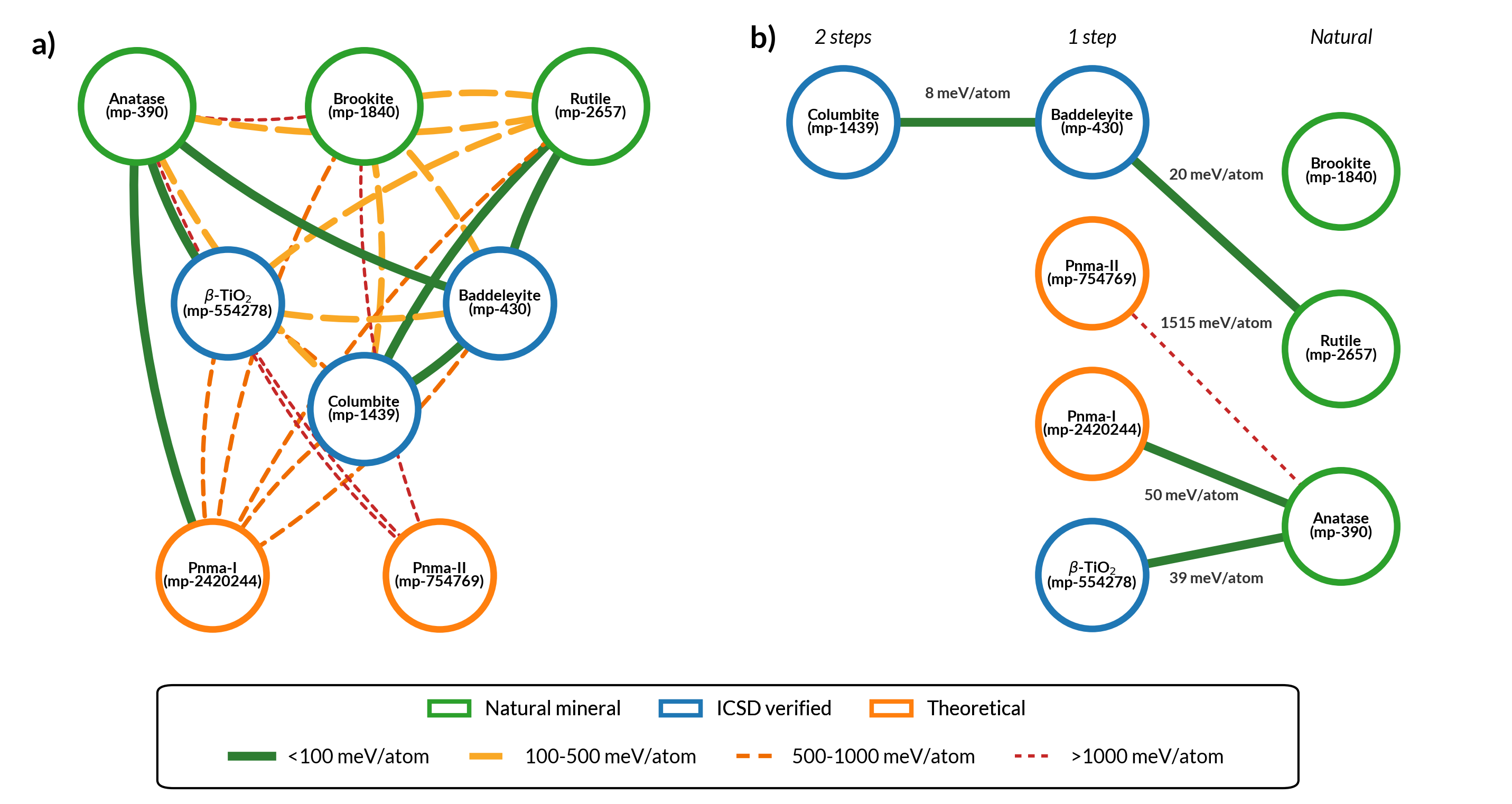}
    \caption{\textbf{Qualitative pathway difficulty between polymorphs.} The weight, color, and style of the edges connecting nodes reflect the height of the energy barrier between the phases they connect. a) All pathways identified between polymorphs b) depiction of the lowest energy pathways separating polymorphs from the naturally occurring phases. In the case of columbite, the lowest energy pathway is a 2-step pathway.}
    \label{fig:phase_graph}
\end{figure}

The overall kinetic trapping effect of this network of pathways is shown in Figure \ref{fig:phase_graph}. In Figure \ref{fig:phase_graph}b, we show the lowest single pathways leading from the high-pressure or hypothetical phases to the naturally occurring phases. We find facile pathways (<100meV/atom - for columbite, this is a two step transformation) for all phases but Pnma-II (mp-754769, theoretical) which is only connected to brookite (mp-1840) via a path with a very high energy barrier. These results supply more explanation for the absence of the phases in nature, namely that there is no large energetic barrier preventing them from transforming into one of the lower energy, naturally occurring phases.

\begin{figure}
    \centering
    \includegraphics[width=0.8\linewidth]{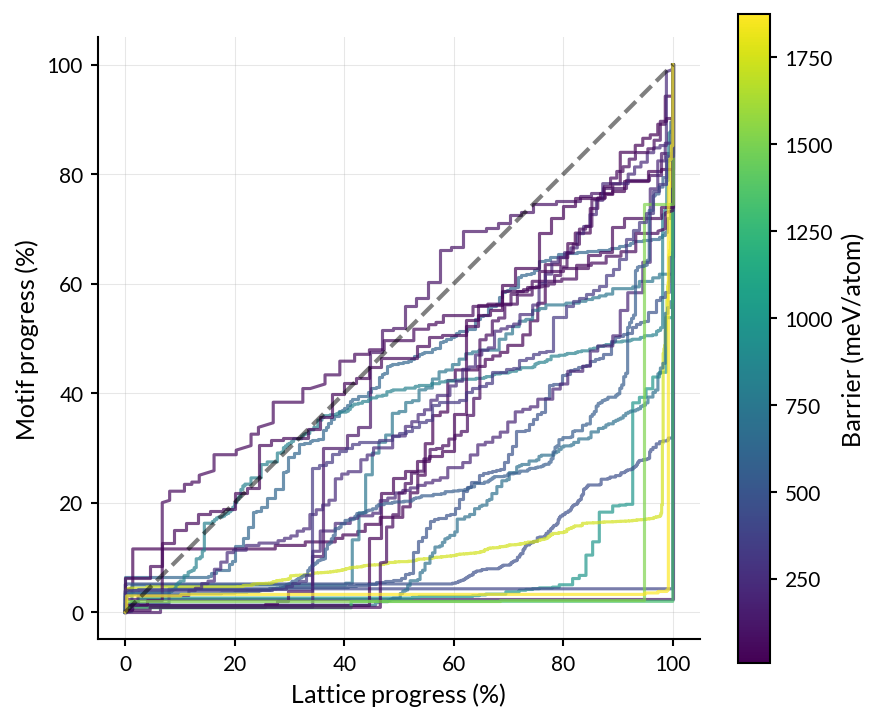}
    \caption{\textbf{Pathway character and energy barrier height.} One trace is shown for each polymorph pair for which a connecting path was identified. The height of the energy barrier associated with each path is given by the trace color. The percentage of motif and lattice progress is calculated by counting the number of steps in each category along each path.}
    \label{fig:pathway_character}
\end{figure}

A distinguishing feature of this pathfinding algorithm is that it produces paths composed of complex mixes of strains and atomic shuffles. It is straightforward to categorize each step as either a ``lattice step'' (a small strain) or a ``motif step'' (a small atomic shuffle). It is then possible to count the total number of steps of each type and measure the progress along each dimension over the course of the transformation. We illustrate this in Figure \ref{fig:pathway_character}. Paths which stay near the diagonal have more evenly dispersed motif and lattice steps while paths that stay near the x-axis begin with a distinct period of unit cell deformation and end with a distinct period of atomic shuffle. Lower-energy pathways cluster near the diagonal, indicating that pathways with more balanced, concurrent strain and shuffle components are energetically favored.

Finally, we note that all pathfinding attempts between Pnma-II (mp-7547687, theoretical) and other polymorphs either failed to produce any path that remained below a 2000 meV/atom or found only high energy paths (>1500 meV/atom). Additionally, our method fails to identify a low or moderate energy pathway between brookite and anatase (we find a 1316 meV/atom barrier height). This result is inconsistent with previous work which recovered low energy pathways, though those previous studies d are not in strong agreement regarding the height of those barriers: Chen et al. found a pathway with a barrier height of 73 meV/atom \cite{chen_phase-transition_2022} and \cite{song_phase_2020} found one with a barrier height of 320 meV/atom. A potential explanation for the tendency of the algorithm to find high energy pathways for these cases lies in the size of the unit cells. Because the unit cells of brookite and Pnma-II (mp-754769, theoretical) contain 24 atoms, the search process is inherently more complex and thus more likely to find only high energy pathways. Higher energy pathways require less compute time to identify than lower energy pathways because a larger region of the CNF graph is available for exploration when the energy ceiling is higher. Specifically, more nodes are below the energy ceiling and thus included in the search.

Additionally, we highlight that the high energy pathways we identified involving 24-atom unit cells (all paths involving Pnma-II, and the brookite$\leftrightarrow$Pnma-I, brookite$\leftrightarrow$anatase, and brookite$\leftrightarrow\beta$-\ce{TiO2} paths) tend to exhibit a 2-stage character: a period of lattice distortion is followed by a period of motif distortion (Figure \ref{fig:pathway_character}). It is possible that this tendency is an artifact of the combination of our Manhattan distance heuristic, which guides the A* shortest pathfinding algorithm, and the long CNF coordinate strings associated with 24-atom unit cells. Because the CNF representation employs a canonicalization procedure to achieve structure disambiguation, the coordinates describing the motif may be reordered during a pathfinding step when the lattice changes. As a result, early alignment of motif coordinates with the target motif coordinates can be disrupted by subsequent lattice changes, which trigger motif coordinate reordering and may increase the apparent distance to the goal. This effect becomes more pronounced for larger unit cells, as the number of motif coordinates, and thus their contribution to the heuristic distance, scales with system size ($3 * (n_{atoms} - 1)$), while the number of lattice coordinates remains fixed at seven. Consequently, the failure to identify the low-energy pathways highlighted in the literature \cite{chen_phase-transition_2022} between these structures may reflect a heuristic bias that rewards lattice alignment before motif alignment. 

Ultimately, we believe that this heuristic bias effect and the overall failure of this algorithm to identify the previously reported low-energy transformation paths for brookite are results of the massive size and complexity of the discretized configuration space represented by the CNF graph for larger unit-cells. This occurs because the paths are long (>1000 steps for discretization parameters that allow realistic configurations to be explored), the branching factor is extremely high (>120 neighbors per node), and the heuristic provides no distinguishing information for nodes in the search frontier, causing the A* algorithm to decay to uniform search. That said, the algorithm is guaranteed to find a path given enough computational time, even at lower energy ceilings, so long as one exists. In light of this, while we elected to use a 24hr cutoff and a procedurally determined maximum number of iterations in the A* pathfinding algorithm efforts in this study, future valuable work will entail modifications of the A* algorithm used here, adjustments of the pathfinding parameters, and experiments using increased computational budgets.

\section{Methods}

\subsection{MLIP Finetuning}

\begin{figure}[h]
    \centering
    \includegraphics[width=0.6\textwidth]{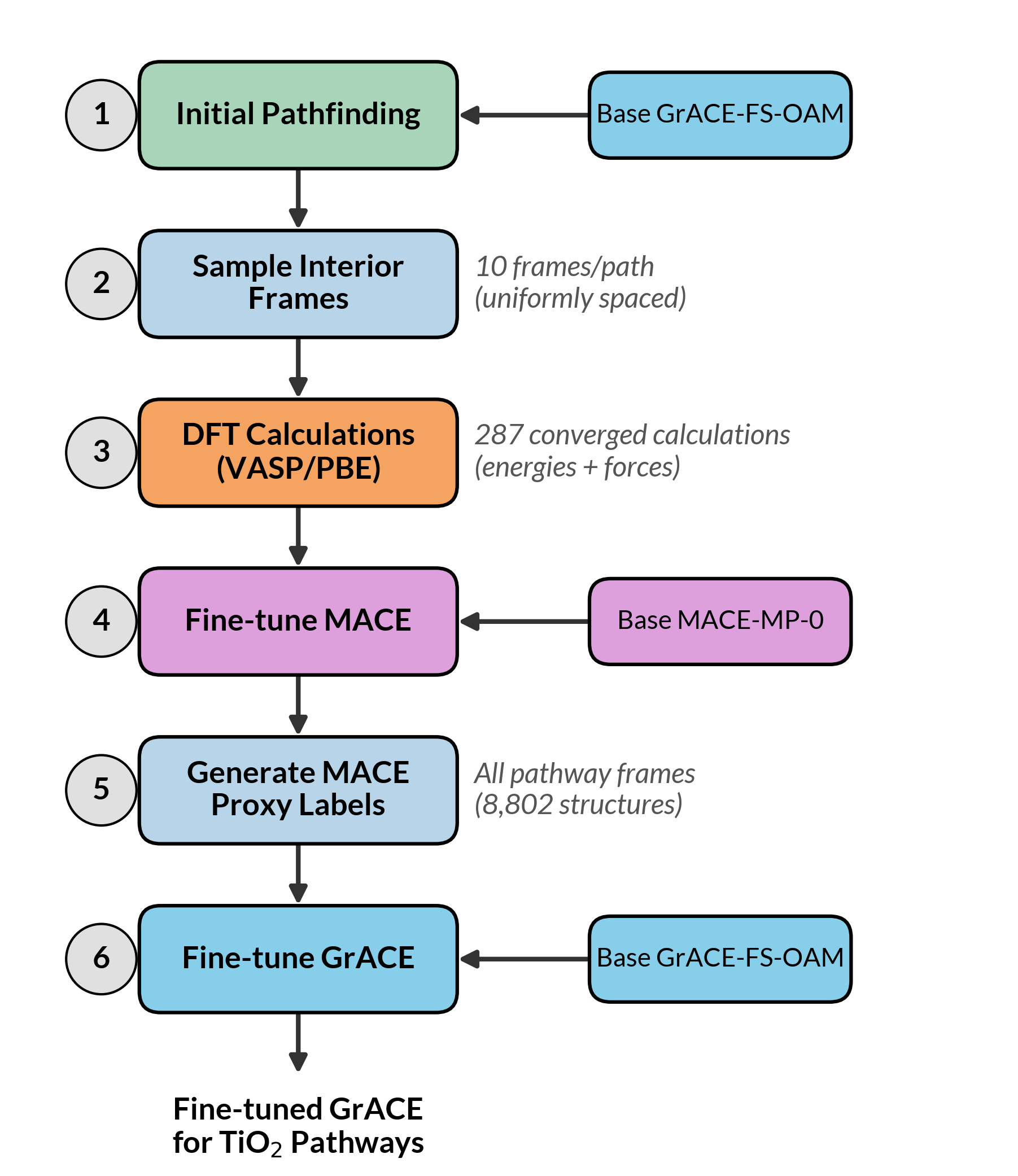}
    \caption{\textbf{Description of the finetuning and distillation process.} Initially, a GrACE pretrained model is used to generate pathways between polymorphs. From the lowest energy pathways found, frames are sampled and their energies are computed to create train and test sets for finetuning. A larger MACE model is finetuned on these frames and that finetuned MACE model is then distilled by finetuning the pretrained GrACE model on a large training set consisting of thousands of frames (and energies predicted by the finetuned MACE model) selected from the initial pathway search.}
    \label{fig:finetune_wf}
\end{figure}

Because millions of energy evaluations are required for these pathfinding efforts, a fast energy evaluator is required. DFT calculations are orders of magnitude too slow (minutes to ~1 hr using GGA for configurations up to 24 atoms). Furthermore, we found that graph neural network MLIPs such as the message-passing atomic cluster expansion (MACE) \cite{batatia_foundation_2025} were 5-10 times slower than linear graph atomic cluster expansion (GrACE) \cite{lysogorskiy_graph_2026} models. As a result, we elected to distill a GrACE model from a finetuned MACE model which was itself fine-tuned on a dataset DFT data for configurations taken from the central regions of a first pass of transformation pathways identified using our method. This workflow is illustrated in Figure \ref{fig:finetune_wf}.

\begin{figure}[h]
    \centering
    \includegraphics[width=\linewidth]{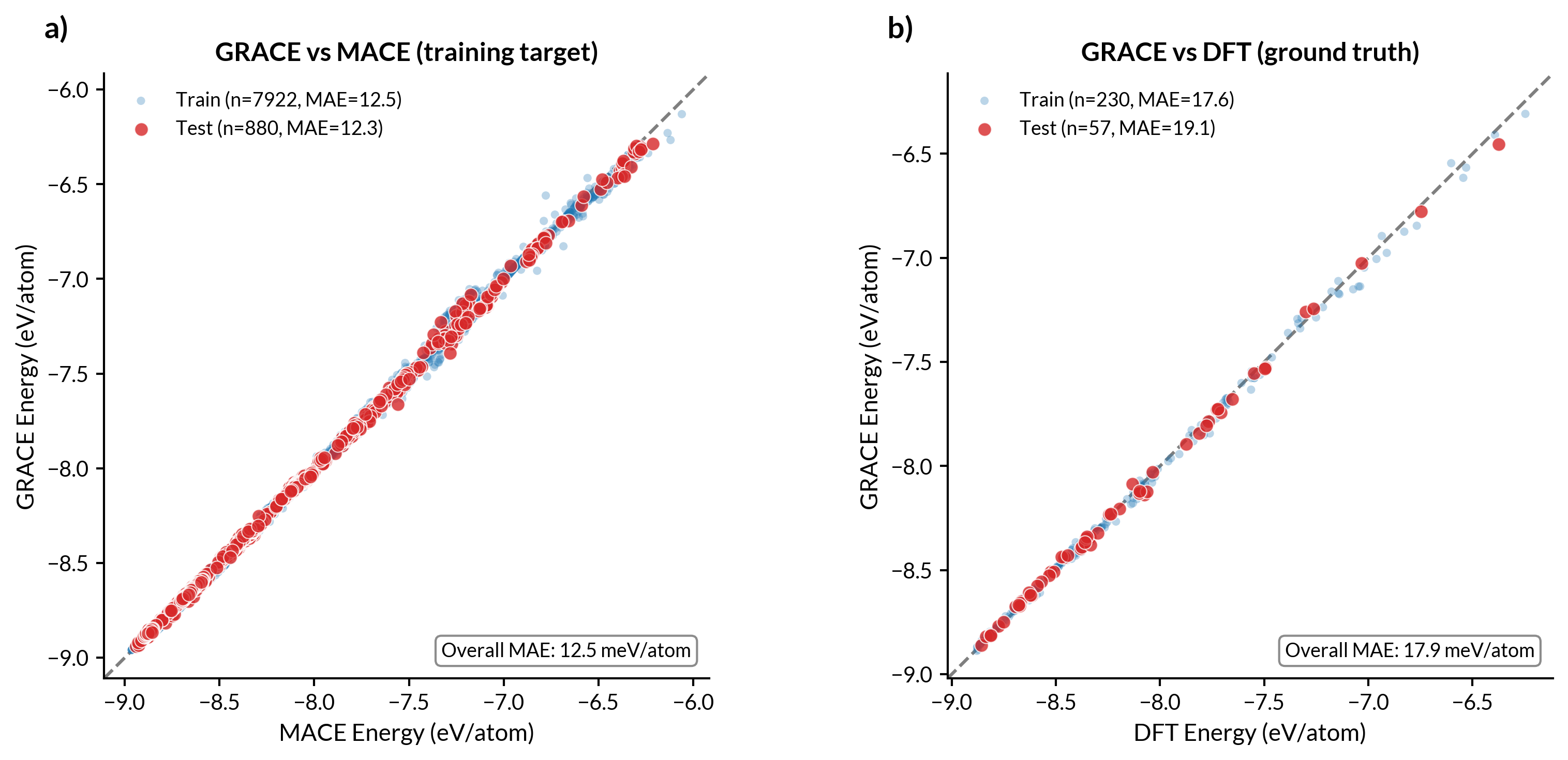}
    \caption{\textbf{Parity plots for the finetuned GrACE potential.} Parity plots for the GrACE model with respect to a) the dataset used to train it (generated by the DFT-finetuned MACE model), and the corresponding MACE-generated test set and b) the DFT-dataset used to finetune the MACE model before distillation. The GrACE model was never directly trained on this DFT data but by distillation still achieves a low 19.1 meV/atom accuracy on the test set.}
    \label{fig:grace_parity}
\end{figure}

To produce training data for this workflow, we executed the pathfinding algorithm on every pair of \ce{TiO2} polymorphs enumerated in Table \ref{tab:polymorphs} and utilized a pretrained GRACE-FS-OAM model as the energy evaluator. From each of the resulting successful paths, we sampled 10 frames from the central region of the path. We computed the energies of those configurations by using DFT calculation via the Vienna Ab Initio Simulation Package (VASP) parameterized by the MPGGAStaticMaker\cite{ong_python_2013, ganose_atomate2_2024} \verb|atomate2| workflow which uses the generalized gradient approximation (GGA) functional of Perdew, Burke and Ernzerhof \cite{perdew_generalized_1996}. We separated the resulting dataset into a 90/10 train/test split and fine-tuned the MACE-MP-0 (small) foundation model using the training portion of this dataset. Using the resulting model, we produced a large training set consisting of every frame from the initial pathfinding procedure, and fine-tuned GRACE-FS-OAM on this dataset, effectively distilling the finetuned MACE model into a linear GrACE model. The finetuned MACE model achieved a high accuracy of 12 meV/atom on the held out test set. The distillation process improved the accuracy of the pretrained GrACE model on the same DFT test set from 184 meV/atom (pretrained) to 19.1 meV/atom (finetuned). The parity plots vs. the MACE-generated training set and the ground truth DFT test set are shown in Figure \ref{fig:grace_parity}.

\subsection{Application of the pathfinding algorithm}

To identify pathways between each pair of polymorphs from Table \ref{tab:polymorphs}, we performed a parallel search procedure. Specifically, we selected a range of 32 ceiling energies evenly spaced between 0 meV/atom and 2000 meV/atom, and assigned them to each of 32 worker processes, each performing its own refinement beginning at its own ceiling. We used a uniform lattice step size of $\xi = 0.75$ \AA$^2$ and selected $\delta$ values such that the maximum atom step size with respect to the initial structure was 0.1 \AA. We note that because the largest atom step size is given by the largest lattice parameter divided by $\delta$, the size of an atom step changes throughout the pathway, as the lattice changes. Nevertheless, starting with an initial value of $\delta$ corresponding to a small atom step leads to the real step size remaining small throughout the pathfinding procedure. The parallel worker process proceed in parallel, each iteratively reducing their own ceilings as pathways are found. The trajectory of these ceiling lowerings are shown in Figure S1. We report the lowest energy barrier pathway found for each pathway in Figure \ref{fig:barrier_matrix}. We allowed each of these processes to run for 24 hours. To manage the supercomputing resources used for this effort, we leveraged the \verb|jobflow| and \verb|jobflow-remote| Python packages \cite{rosen_jobflow_2024}.

\section{Conclusion}

The algorithm introduced in this study extends the capabilities of the existing CNF path-finding approach and differs from other pathfinding methods in that it requires no prior input on the nature of the transition path. We illustrate its use in the rich \ce{TiO2} system within which we identify many low-energy transformation barriers. Our findings provide plausible mechanistic insight into the lack of experimental realization of five \ce{TiO2} polymorphs.

This algorithm substantially expands the range of systems for which transformation pathway studies could be performed using the CNF, and we believe that it represents only a small portion of the potential utility of the CNF in the discovery of diffusionless transformations and in computational materials science more generally. Future work should entail improvements to this pathfinding method such as unbiased search heuristics, intelligent neighbor pruning functions, fast dataset generation for MLIP training, efficient methods for changing discretization resolution mid-search and improvements to the pathfinding algorithm itself.

\section*{Acknowledgements}

M.C.G. acknowledges support from the Kavli Energy NanoScience Institute (Kavli ENSI) Graduate Student Fellowship. K.P. acknowledges support from the Materials Project program KC23MP, supported by the U.S. Department of Energy, Office of Science, Office of Basic Energy Sciences, Materials Sciences and Engineering Division under contract No. DE-AC02-05-CH11231. This research used resources of the National Energy Research Scientific Computing Center (NERSC), a U.S. Department of Energy Office of Science User Facility located at Lawrence Berkeley National Laboratory, operated under Contract No. DE-AC02-05CH11231. Additional computational resources were provided by the Lawrencium computational cluster at Lawrence Berkeley National Laboratory.

\subsection{Author Contributions}

\textbf{M.C.G.}: conceptualization (lead); software (lead); writing - original draft (lead). \textbf{D.M} conceptualization; software, writing - review and editing. \textbf{K.A.P.}: supervision; funding acquisition; writing - review and editing.

\section{Competing Interests}

All authors declare no financial or non-financial competing interests.

\section*{Supporting information}

The following files are available free of charge.
\begin{itemize}
  \item Individual plots for lattice/motif pathway character; Parity plots for finetuned MLIP; Example of ceiling lowering traces for parallel workers (PDF)
\end{itemize}

\printbibliography

\end{document}


\section{Pathway Character Detail}

\begin{figure}[H]
    \centering
    \includegraphics[width=0.8\linewidth]{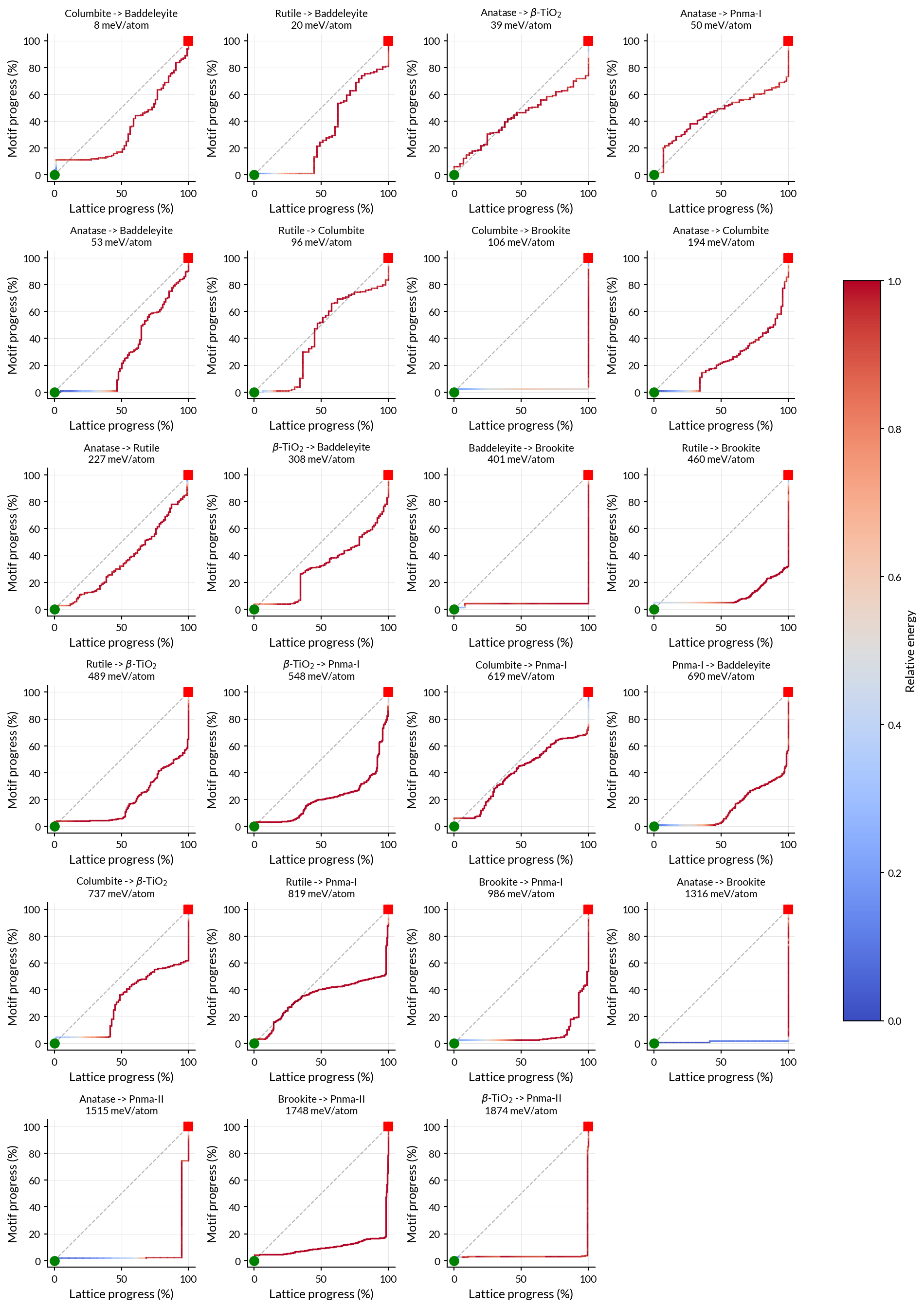}
    \caption{\textbf{Lattice-motif character of each pathway.} Energy color is with respect to the highest energy encountered along the transformation pathway. Pathways with the lowest barrier energy are listed at the top of the figure and pathways with higher barriers are shown at the bottom. The highest energy barrier pathways tend toward distinct two-step character: first a period of lattice alignment then a period of motif alignment.}
    \label{fig:barrier_matrix}
\end{figure}

\section{Parallel refinement result illustration}

\begin{figure}[H]
    \centering
    \includegraphics[width=0.95\linewidth]{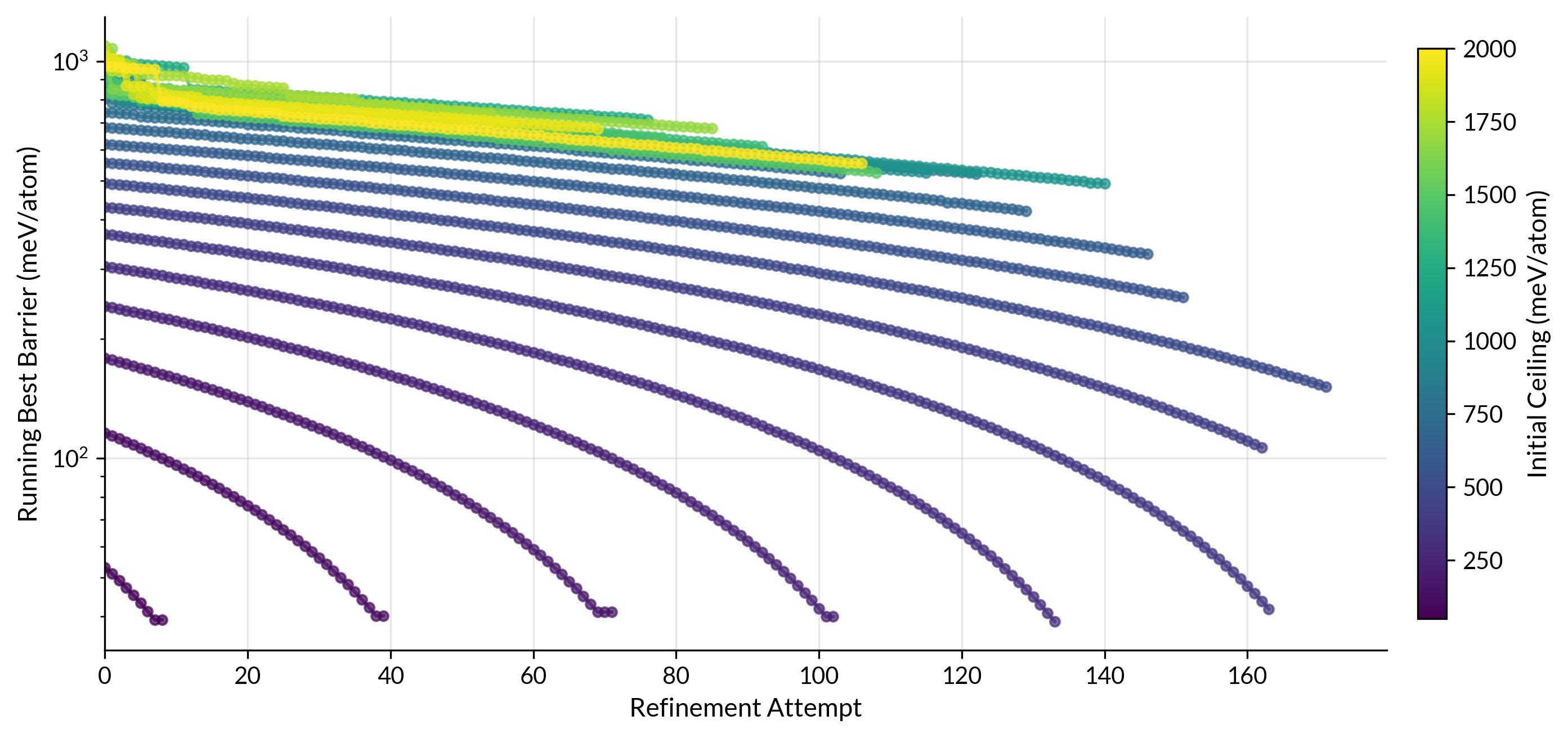}
    \caption{\textbf{Illustration of parallel workers refining ceiling level.} Each trace corresponds to a worker refining a specific ceiling level. Workers start their refinement processes over a range of initial energy ceilings so that the true ceiling level can be quickly found. In this example, the true barrier height is found quickly by the lowest energy worker and then later by several other workers..}
    \label{fig:barrier_matrix}
\end{figure}

\section{GrACE MLIP Verification}

\begin{figure}[H]
    \centering
    \includegraphics[width=0.95\linewidth]{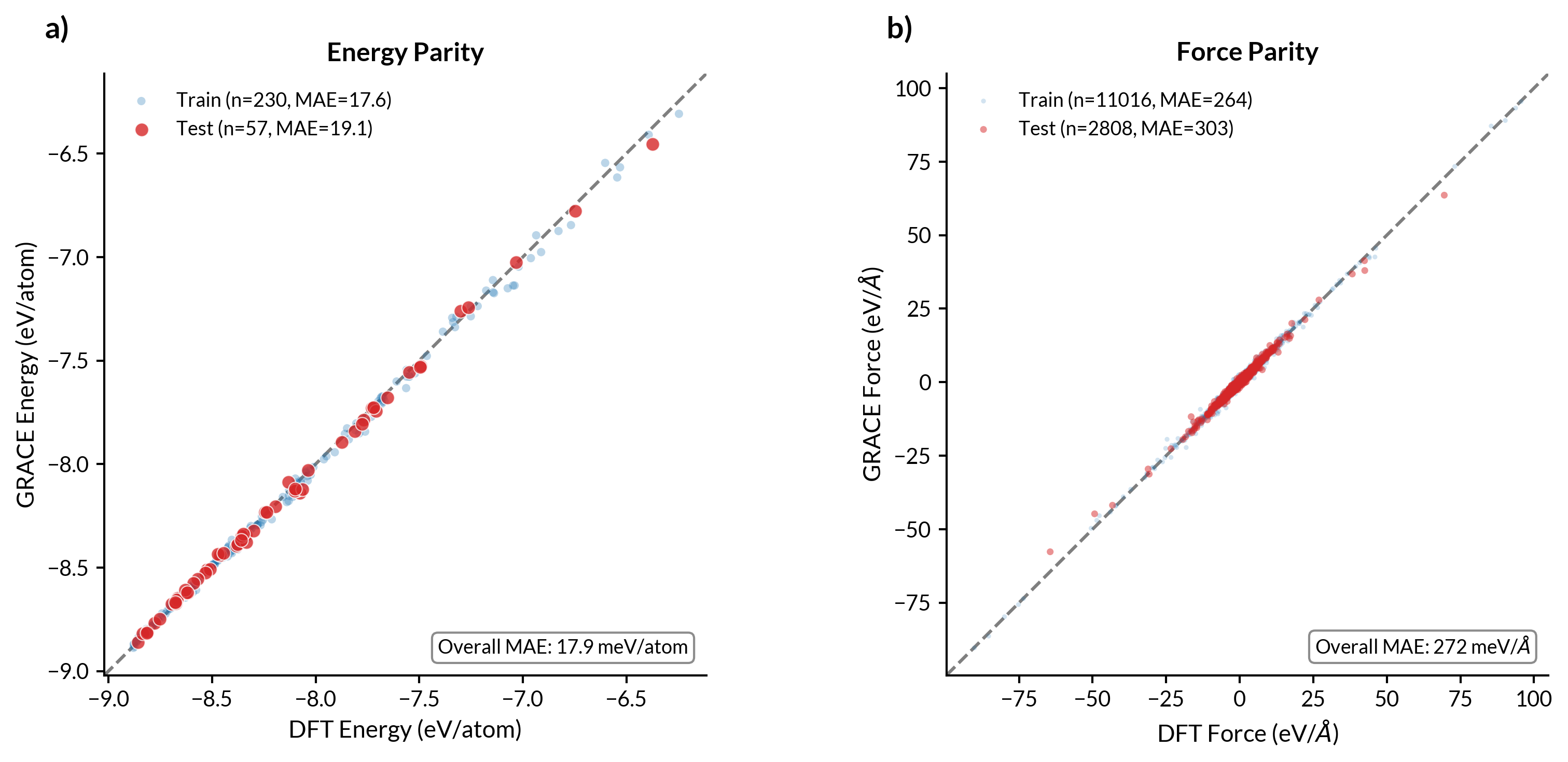}
    \caption{\textbf{GrACE MLIP benchmarking plots.} a) Energy parity plot with respect to the DFT training set used to finetune the teacher MACE model. and b) force parity plot showing overall error with respect to forces computed for the large finetuning dataset produced by the teacher MACE model.}
    \label{fig:barrier_matrix}
\end{figure}